\documentclass[12pt]{article}
  \usepackage[round,authoryear]{natbib}

 \usepackage[latin1]{inputenc}
 \usepackage{float}

 \usepackage{authblk} % A good solution in case of multiple authors having different affiliations

 \usepackage{amsfonts}
 \usepackage{amsthm}
  \usepackage{amssymb, latexsym, amsmath}
 \usepackage{amsbsy}
 \usepackage{amstext}
 \usepackage{amsgen}

 \usepackage{graphicx}
 \usepackage[normalem]{ulem}
 \usepackage[usenames, dvipsnames]{color}
 \usepackage{wasysym}
 \usepackage{verbatim}
 \usepackage{color}
 \usepackage[T1]{fontenc}
 \usepackage[scale=0.8]{geometry}
\usepackage{fancyhdr}
\usepackage{bm}

 \newcommand{\s}{\nobreak\hspace{.11em}\nobreak}

 \newcommand{\be}{\begin{equation}}
 \newcommand{\ee}{\end{equation}}
 \newcommand{\ba}{\begin{eqnarray}}
 \newcommand{\ea}{\end{eqnarray}}
 \newcommand{\bs}{\begin{subequations}}
 \newcommand{\es}{\end{subequations}}

\begin{document}
   \title{
 {\Large{\textbf{{{Initial Conditions for Tidal Synchronisation\\ of a Planet by Its Moon}
 ~\\~\\}
            }}}}

\author{
                                          {\Large{Valeri V. Makarov}}\\
                                          {\small{US Naval Observatory, Washington DC 20392 USA}}\\
                                          {\small{valeri.makarov$\,$@$\,$gmail.com
                                             }}
   \vspace{3mm}
   ~\\
 %                                         {\Large{Ciprian Berghea}}\\
 %                                         {\small{US Naval Observatory, Washington DC 20392 USA}}\\
 %                                         {\small{ciprian.t.berghea$\,$@$\,$navy.mil
 %                                            }}
 %
 %  \vspace{3mm}
 %  ~\\
 {{and}}
 \vspace{4mm}
 ~\\
                                          {\Large{Michael Efroimsky}}\\
                                          {\small{US Naval Observatory, Washington DC 20392 USA}}\\
                                          {\small{michael.efroimsky$\,$@$\,$gmail.com
                                             }}
  }

     \date{}

 \maketitle

\abstract{
Moons tidally interact with their host planets and stars. A close moon is quickly synchronised by the planet, or becomes captured in a higher spin-orbit resonance. However, the planet requires much more time to significantly alter its rotation rate under the influence of moon-generated tides. The situation becomes more complex for close-in planets, as star-generated tides come into play and compete with the moon-generated tides.
Synchronisation of the planet by its moon changes the tidal dynamics of the entire star-planet-moon system and can lead to long-term stable configurations. In this paper, we demonstrate that a certain initial condition must be met for this to occur.
Based on the angular-momentum conservation, the derived condition is universal and bears no dependence upon the planet's internal structure or tidal dissipation model. It is applicable to dwindling systems as well as tidally expanding orbits, and to the cases of initially retrograde motion.
We present calculations for specific planet-moon systems (Earth and the Moon; Neptune and Triton; Venus and its hypothetical presently-extinct moon Neith; Mars, Phobos, and Deimos; Pluto and Charon), to constrain the dynamically plausible formation and evolution scenarios. Among other things, our analysis prompts the question of whether Pluto and Charon evolved into their current state from an initially more compact configuration (as is commonly assumed) or from a wider orbit---a topic to be discussed at length elsewhere.
Our results are equally applicable to exoplanets. For example, if asynchronous close-in exoplanets are detected, the possibility of tidal synchronisation by an exomoon should be considered.}

\section{Introduction}
\label{Introduction}

There are two pathways for tidally interacting planet--moon systems to achieve complete mutual synchronisation, defined as an equilibrium state where both partners' angular rotation rates equal the orbital mean motion.

One path is for secularly expanding pairs, such as the Earth--Moon system today. For this expansion to start, the planet's rotation must initially be faster than the orbital motion of the moon; equivalently, the initial separation of the bodies must be larger than the synchronous radius. Also, the initial eccentricity should not be too high~\citep{2021NatAs...5..539B, 1983Icar...55..133S}. This initial setting ensures that the principal tidal bulge raised on the planet by the moon is aligned (up to libration) with the instantaneous direction toward the moon, and that the planet is despinning, allowing for the angular momentum to be transferred from the planet's rotation to its orbit.

The other path is for initially closer satellites with orbital periods shorter than the planet's rotation period. The orbital momentum is then transferred from the orbit to the planet's rotation due to the torque from the {lagging} tidal bulge on the planet. The best known example of this option is the Mars--Phobos pair. {The case of a retrograde moon, such as Triton~\citep{2009AJ....137.4322J} or the hypothetical satellite of Venus~\citep{2023Univ...10...15M}, also belongs in this category.}

For close-in planets with moons, the significant presence of the star--planet interaction adds layers of complexity to these scenarios, providing either accelerating or counter-directed tidal action on the planet~\citep{2023A&A...672A..78M}.

When the moon is sufficiently massive and close to the planet, the tidal action from it overwhelms the torque on the planet from the star. Two outcomes of secular tidal evolution are then possible: in one, the moon crosses the Roche limit; in the other, the planet's rotation rate evolves toward an equilibrium state, a spin--orbit resonance with the~moon.

This topic was first addressed in the pioneering work by~\citet{Darwin1879}, who found that, for a conserved angular momentum of a two-body tidal system, at most two synchronous configurations are possible, with the one of lower energy being stable. A century later, the problem was addressed again by~\citet{Hut_1980}, who investigated {a primary mass rotating at a rate $\Omega$ and a point-like secondary mass} describing an orbit with a mean motion $n$, inclination $i$, and eccentricity $e$. He demonstrated that, rigorously speaking, the only tidally stable spin--orbit state is the one satisfying the conditions of circularity ($e=0$), coplanarity ($i=0$), and corotation ($n=\Omega$). Conditions of stability were obtained via eigenvalues of the energy Hessian in the space of total angular momentum and $n$ for an idealised two-body system.

Our study is a development of this problem, aimed at defining the initial conditions of the observable parameters (the semimajor axis $a$ and the planet's rotation rate $\Omega$) that can result in configurations remaining transiently stable for sufficiently long periods in realistic three-body systems. We derive analytical inequations and specify the dualism of synchronous configurations (long-term stable versus inherently unstable) from first principles, irrespective of the specific tidal model. Our results can be directly applied to a variety of solar or exoplanet configurations, as illustrated by several examples.

Within realistic tidal models for terrestrial bodies~\citep{2012CeMDA.112..283E}, higher-order spin--orbit resonances can emerge at 3:2, 2:1, and even higher commensurabilities~\citep{2014Icar..241...26N}. This can occur, probabilistically, even if the amplitude of the tidal torque in the vicinity of a given resonance is somewhat smaller than the secular torque caused by the finite eccentricity and the triaxial shape of the moon. Assuming the current eccentricity value, the moon was more likely to be captured in a 3:2 spin--orbit state than to miss it during the course of tidal despinning~\citep{1966AJ.....71....1G, 2013MNRAS.434L..21M}. A higher spin state, however, is only a local equilibrium, because a sufficiently strong external perturbation (e.g., an impact) or a partial meltdown of the moon's crust~\citep{2018arXiv180405110H} can drive the moon out of this state.
% Therefore, we only address planet--moon configurations with a moon spinning synchronously (or pseudosynchronously, if the moon is still hot and has a low viscosity), because, in such a state, the tidal action from the moon achieves the global minimum. We also assume the eccentricity to be small and neglect the spin angular momentum of the moon.

Our goal is to adumbrate a set of initial conditions that must be met by a tidal binary to make tidal synchronism attainable. These conditions are independent of the planet's tidal properties and rheology. We present calculations for specific planet--moon systems (Earth and the Moon, Pluto and Charon, Mars and its satellites, and Venus and its hypothetical presently extinct moon Neith) to constrain dynamically plausible formation and evolution scenarios.

\section{Angular momentum exchange}
\label{con.sec}

This discussion expands upon the ideas proposed by~\citet{2023A&A...672A..78M}.

We consider a planet of mass $M$ and radius $R$ and a moon of mass $M_m$ and radius $R_m$. Their maximal moments of inertia are $\xi M R^{\s 2}$ and $\xi_m M_m R_m^{\s 2}\s$, where $\xi$ and $\xi_m$ are dimensionless moment of inertia (MOI) coefficients. As customary, $a$ and $n$ are the semimajor axis and mean motion.

{For mathematical convenience, we shall always assume the mean motion to be positive, $n > 0$, while the rotation rate $\Omega_p$ of the planet will be positive or negative, depending on whether the moon's orbit is prograde or retrograde.}

To single out the principal aspects of the dynamics, three simplifications are adopted:
\begin{itemize}
    \item[(a)] The moon's spin angular momentum can be neglected.
\end{itemize}
This premise is reasonable, even when the partners' masses differ by less than an order of magnitude. For example, the spin angular momentum of Charon is about $0.04\%$ of the orbital angular momentum of the system and about $3\%$ of the spin angular momentum of~Pluto.

\begin{itemize}
    \item[(b)] The moon's orbit is near-circular, and $O(e^2)$ terms may be omitted.
\end{itemize}
We would add that the evolution of $e$ is a nontrivial issue, as the direction of this evolution is sensitive to the partners' rheologies and, therefore, not immediately apparent.\,\footnote{~Within the Constant Phase Lag (CPL) model (i.e., for a frequency-independent $k_2/Q$), the tides in both the synchronised and nonsynchronised partners are working to increase the eccentricity. Within the Constant Time Lag (CTL) model (i.e., for $k_2/Q$ linear in the tidal frequency), tides in a body mitigate $e$ when its spin rate divided by the mean motion exceeds $18/11$, and are boosted $e$ otherwise. For more realistic rheologies, the situation becomes more complex (see, e.g., a short summary on this in Appendix E in {\cite{Mars}.)}}
 ~However, even when $e$ is predicted to approach zero in a simplified two-body model with tides included, a small but finite value of $e$ may persist in a real-world scenario due to the influence of external bodies and/or a resonance between the orbital motion and the irregular shape of the host planet. Despite these complexities, we take the liberty of assuming that $e$ is small and neglects the terms of order $e^2$.

\begin{itemize}
    \item[(c)] The moon's orbital inclination $i$ relative to the planet's equator is small, as is the moon's obliquity $i_m$ relative to its orbit. Therefore, all terms of order $O(i^2)$ or $O(i_m^2)$ may be dropped.
\end{itemize}
Neglecting the spin angular momentum of the moon, and ignoring the $O(e^2)$, $O(i^2)$, and $O(i_m^2)$ terms, the conservation of angular momentum reads as follows (Appendix D in~\citet{2023A&A...672A..78M}):
\be
\frac
{{\stackrel{\bf\centerdot}{\Omega}}_p}
{{\stackrel{\bf\centerdot}{n}}}
\,=\,\frac{1}{3\,\xi}\,\frac{M_m}{M+ M_m}
\left(
\frac{a}{R}
\right)^2+\,O(e^2)\;\;,
\label{3ksi.eq}
\ee
where ${\stackrel{\bf\centerdot}{\Omega}}$ is the planet's rotation rate. Note that this equation holds for positive or negative time derivatives in the left-hand part. As was explained in Section~\ref{Introduction}, there are two distinct evolutionary paths characterised by either both $\dot n$ and $\dot \Omega_p$ positive (implying orbital decay and the planet's spin-up) or both derivatives negative (implying orbital expansion and the planet's spin-down). Implicitly, the semimajor axis $a$ is a function of time.

In Equation~(\ref{3ksi.eq}), it is convenient to remove ${\stackrel{\bf\centerdot}{n}}$ in favour of ${\stackrel{\bf\centerdot}{a}}$. From $\,n\propto a^{-3/2}\s$ follows $\;{\stackrel{\bf\centerdot}{n}}\s=\,-\,\frac{\textstyle 3}{\textstyle 2}\,a^{-1}\s n\,{\stackrel{\bf\centerdot}{a}}\,$, the insertion whereof into Formula~(\ref{3ksi.eq}) entails
\be
\dot\Omega_p = -\frac{X}{2}\,\frac{\dot a}{\sqrt{a}\,}\;\;,
\label{dotO.eq}
\ee
where the dimensional factor $X$ defined through
\ba
X=\frac{M_m}{\xi}\,\sqrt{\frac{G}{M_{\rm tot}}\s}\s R^{\s -2}
\;\;,
\qquad
M_{\rm tot}= M+ M_m
\;\,,
\label{X}
\label{4}
\ea
may be interpreted as a measure of the angular momentum exchange rate.

Introducing the combination
\ba
C=\Omega_p(t_0)+X\sqrt{a(t_0)}\;\,,
\label{C}
\ea
of the initial conditions, we observe that a direct integration of Equation~(\ref{dotO.eq}) over time produces
\ba
\Omega_p=-X\sqrt{a}+C\,\;,
\label{5}
\label{Omega}
\ea
where $C$ appears as an integration constant. Therefore, the planet's rate of rotation $\Omega_p$ is a linear function of $\sqrt{a}$ with a negative slope.

In these variables, the synchronisation condition $\Omega_s=n$ translates to a cubical hyperbola in terms of $\sqrt{a\s}\,$:
\ba
\Omega_s=\sqrt{G\,M_{\rm tot}\,}\,(\!\sqrt{a})^{-3}\;.
\label{6}
\label{OmegaS}
\ea

\begin{figure}[H]
    \includegraphics[width=0.6\textwidth]{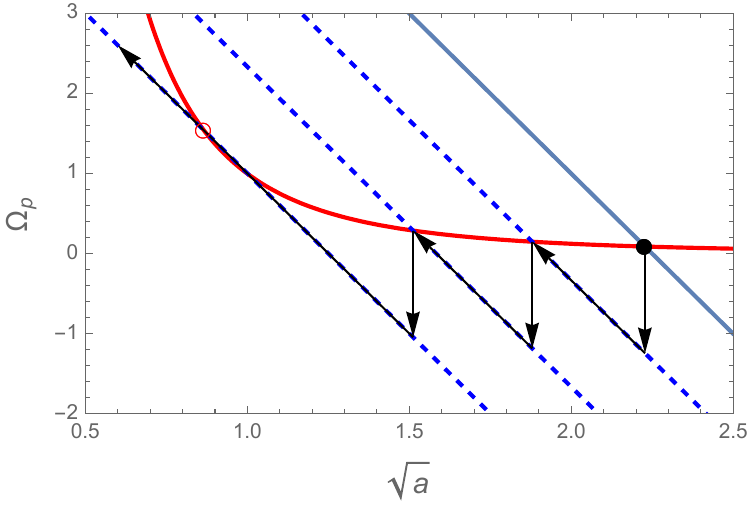}
    \caption{{Tidal evolution}
    %MDPI: Please add the explanation for red circle and solid/dashed lines in the figure.
 of a toy planet--moon system in the $(\sqrt{a}\s,\,\Omega_p)$ axes. As follows from Equations~(\ref{Omega}) and (\ref{OmegaS}), straight lines with a negative slope are evolutionary tracks, while the cubical curve represents all available synchronous states. The direction of evolution along the straight lines is shown with inclined arrows. Chosen as initial, the state marked with a filled black circle is that of the stable equilibrium. For example, consider a series of impulsive perturbations from the star (tidal deceleration of the planet's rotation), shown as downward black arrows. Each such push entails a subsequent action from the tides raised by the moon on the planet. Depicted with arrows pointing to the left and up, these actions return the system to new states of equilibrium. The lowest linear track passes through a critical tangent point marked with a red circle, where the planet cannot be secularly synchronised any more.}
    \label{toy.fig}
\end{figure}

Thus, we obtain two functions, shown in Figure~\ref{toy.fig}.

The function $\Omega_p(\!\sqrt{a})$, given by Equation~(\ref{5}) and depicted by a blue straight line, represents angular momentum conservation, a condition that must be satisfied by each physical history. Different straight lines in the figure correspond to different values of the angular momentum carried by the system. The location and slope of each such line depends, via $X$, on the parameters of the system, and, via $C$, on the initial conditions.

The function $\Omega_s(\sqrt{a})$, given by Equation~(\ref{6}) and shown by a red cubical hyperbola in the figure, represents the planet's angular velocity synchronised to the orbital motion with a semimajor axis $a\s$. This curve is only defined by the total mass, not by the initial~conditions.

Not every physical history (i.e., not every possible straight line with a fixed slope in Figure~\ref{toy.fig}) can intersect the red curve. In other words, not every set of the parameter values and initial conditions sends the system towards synchronism. When the initial conditions and parameter values are such that curves (\ref{5}) and (\ref{6}) do not intersect, this means, physically, that the planet's rotation rate cannot catch up with the mean motion on the contracting orbit, and the system cannot be synchronised.

{The situation is borderline when} lines (\ref{5}) and (\ref{6}) are tangent in a single point. This brings up two conditional equations:
\ba
\left.
\begin{array}{lll}
\frac{\textstyle \partial \Omega_p}{\textstyle \partial \sqrt{a}}\,=\,\frac{\textstyle \partial \Omega_s}{\textstyle \partial \sqrt{a}}
~\\
~\\
\Omega_p(\!\sqrt{a})= \Omega_s(\!\sqrt{a})\!\!
\end{array}
\right\}
\Rightarrow
\left\{
\begin{array}{lll}
\!\!-X\,=\,-3\,\sqrt{G\,M_{\rm tot}}\s(\!\sqrt{a})^{-4}
~\\
~\\
\!\!-X\sqrt{a}+C_{\rm syn}=\sqrt{GM_{\rm tot}}\s(\!\sqrt{a})^{-3}\;,
\end{array}
\right. \!\!\!\!\!\!
\label{7}
\ea
where the constant $C_{\rm syn}$ is the specific value of $C$ corresponding to the borderline case. Formula (\ref{7}) renders
\bs
\ba
C_{\rm syn}=4\s\left(\frac{X}{3}\right)^{3/4}\!\left(G\,M_{\rm tot}\right)^{1/8}\;\,,
\ea
which, in combination with expression (\ref{X}), entails
\ba
C_{\rm syn}=\frac{4}{(3\,\xi)^{3/4}}\frac{\sqrt{G}\,M_m^{3/4}}{M_{\rm tot}^{1/4}}\, R^{-{3/2}}\;\,.
\ea
\label{Csyn}
\es

Synchronisation is possible if the actual constant of integration $C$ exceeds or equals the borderline value $C_{\rm syn}\,$:
\be
C\geq C_{\rm syn}
~~~\Longleftrightarrow~~~
\Omega_p(t_0)+X\sqrt{a(t_0)}\s\geq\s
4 X \sqrt{R}\left(\frac{\xi}{27}\,\frac{M_{\rm tot}}{M_{ m}}\right)^{1/4}
\;,
\label{cond.eq}
\ee
where we invoked definitions (\ref{X}), (\ref{C}), and (\ref{Csyn}).

The condition (\ref{cond.eq}) of synchronism attainability depends on the parameter values and the initial conditions $\Omega_p(t_0)$ and $a(t_0)$. It is valid for any scenario of tidal evolution, including the case of an initially retrograde rotation ($\Omega_p(t_0)<0$).

In the borderline situation where the parameters and initial conditions are picked such that the weak inequality (\ref{cond.eq}) becomes equality, $C = C_{\rm syn}$, the moon becomes synchronous in a single point $\{\!\sqrt{a}\s,\,\Omega_p\}$. In Figure~\ref{toy.fig}, this situation is illustrated with the point where a straight line is tangent to the red cubical curve. The equilibrium at the tangent point is unstable under a perturbation decelerating the planet's spin (e.g., a tidal torque exerted on the planet by the star) or reducing the moon's orbit. Depicted with downward black arrows, decelerating perturbations are compensated by the action of moon-generated tides (shown with arrows pointing to the left and up). This stabilisation mechanism, however, will not work in the tangent point. We thus observe that the moon can reach the tangent point only by moving %MPDI: Please check that intended meaning is retained.
from right to left, i.e., from an initially wider orbit. As can be seen, the moon will not pause if perturbed, and will keep spiraling down. It is therefore highly unlikely to find a planet--moon system in this transient state.

%To draw this section to a close, we would mention that
The values of the synchronous radii can be found from Equation~(\ref{Omega}) by setting $\Omega_p = n$ in it. In terms of the variable
\ba
s\equiv\sqrt{a}\;\,,
\label{}
\ea
this renders the quartic equation
\ba
X\,s^4\,-\,C\,s^3\,+\,\sqrt{G\,M_{\rm tot}}\,=\,0\;\,,
\label{quartic}
\ea
which has four roots. Three options are available: (a) none of the roots are positive real, meaning that synchronisation is impossible; (b) only one (nonidentical) root is positive real, corresponding to the marginal tangent point condition (\ref{7}); or (c) only two roots are positive real, defining the two synchronous states.

%We note in passing, that
The considerations above are also valid for spin--orbit resonances higher than the synchronous 1:1 state addressed in this paper. Indeed, sufficiently cold, inviscid planets with a moon in an orbit of finite eccentricity can be captured in such a nonsynchronous resonance (e.g., 3:2) if certain conditions are fulfilled on the mass and orbit separation. The only difference would be an additional factor 3/2 in Equation~(\ref{OmegaS}), which would shift the cubical hyperbola in Figure~\ref{toy.fig} upward. This change would set more stringent initial conditions to achieve the resonant state (a higher initial rotation rate $\Omega_p(0)$) and would favour more massive satellites. Generally, a pair of distinct equilibrium states emerges as well, with the inner state being intrinsically unstable.

\section{The stability of synchronism, in local terms}
\label{criterion}

Results that are consistent with the above findings can be obtained using a different method. {Ignoring the tides in the satellite, consider} only the contribution from the tides in the planet into the tidal rate of the semidiurnal axis. In this contribution, {the quadrupole semidiurnal term is}
\ba
\left(\frac{da}{dt}\right)^{\rm (pqs)} \,=\,-\,3\,a\,n
\frac{\;M_m}{M}\,\left(\frac{R}{a}\right)^5\,K_2(2n-2{\Omega}_p)
\label{term3}
\ea
where $K_2(\omega)$ is the quality function and $\omega$ is a short notation for the principal (semidiurnal) tidal Fourier mode $\omega_{2200} = 2n-2\Omega_p\s$. The superscript ``$\rm (pqs)$'' denotes ``$\rm planet, quadrupole, semidiurnal$''. %MDPI: Please confirm if the italics is unnecessary and can be removed. The following highlights are the same.
%Authors: the italics can be replaced by quotation marks

To explore the stability of this term under the small variations of planet--moon separation, we calculate its full (not partial) derivative with respect to $a$ at a point of synchronism where $\s n\s=\s\Omega_p\s$. As demonstrated in Appendix~\ref{Appendix A},
\ba
\frac{d}{da}
\left(\frac{da}{dt}\right)^{\rm (pqs)}_{
\,n= {\Omega\s}_{\!p}
}=
9\frac{\;M_m}{M\,}\left(\frac{R}{a}\right)^5 n^2
\left(1-\frac{d{\Omega}_p}{dn}  \right)\frac{dK_2(\omega)}{d\omega}\s
\Bigg{|}_{\,\omega=0}\;\,\;,
\label{insynchro}
\ea
the derivative ${d{\Omega}_p}/{dn}$ is given by expression (\ref{3ksi.eq}). In Appendix~\ref{Appendix A}, we {explain} that the derivative ${dK_2(\omega)}/{d\omega}$ is positive at the point of spin--orbit resonance crossing for a planet of the Maxwell rheology. This conclusion is valid for all realistic bodies.\,\footnote{~For any rheological model, the positive sign of ${\textstyle dK_2(\omega)}/{\textstyle d\omega}$ at $\omega=0$ is a necessary condition for the synchronous spin--orbit state to be stable.} From this, we deduce a concise and practical criterion of stability:
\begin{itemize}
    \item[] Synchronism is stable if $\s{d\Omega_p}/{dn}>1\s$; unstable otherwise.
\end{itemize}

We term this criterion ``local'' because it bears no explicit dependence on the initial conditions (although it refers to them indirectly, using the assumption that a synchronous state does exist).
The ``local'' criterion, of course, agrees with the ``global'' one derived in the preceding section and illustrated by Figure~\ref{toy.fig}. Indeed, in this figure, stable (unstable) configurations correspond to a situation where the absolute value of the slope $d\Omega_p/d\!\sqrt{a}$ of a straight line exceeds (falls short of) the absolute value of the slope $dn/d\!\sqrt{a}$ of the cubical curve, at the point of the curves' crossing. {In other words, any perturbation of $\Omega_p$ from the inner point of equilibrium, independent of its sign, sends the system on a new linear trajectory with the same direction and the same negative slope, driving the system father away from the red curve in {Figure~\ref{toy.fig},} %MDPI: Figures should be cited in correct numerical order. Please cite Figure 2 before Figure 3 and ensure that figures are placed in the same section as their first citation.
%Authors: corrected by changing the reference to Fig. 1 at this location.
 where it could regain synchronism. For example, a spontaneous or externally incited acceleration of the rotation rate puts the system on a linear tract leading downward and to the right. The opposite situation is observed in the stable outer point of equilibrium, where an upward perturbation in $\Omega_p$ sends the system toward a new and slightly updated state of synchronism. The same consideration is valid for a perturbation of $a$. Thus, the restoring action, which is the necessary condition of stable equilibrium, is only realised in the outer point of intersection. It can be demonstrated that, for a vanishing eccentricity and a neglected spin angular momentum of the secondary, our results, although obtained by a different method, do agree with those derived by~\citet{Darwin1879} and~\citet{Hut_1980}.}

\section{Generalisation to the case of a 3:2 spin-orbit\\
resonance} \label{32.sec}

{While the method of stability analysis described in Section~\ref{criterion} is generic and, specifically,  applicable to an arbitrary spin--orbit resonance, here we shall illustrate how it works in the 3:2 case.
}

Equation (\ref{ae1.eq}) from Appendix~\ref{Appendix A} comprises the quadrupole inputs into $da/dt$ of orders up to $e^4$. To analyse the behaviour of the system in the vicinity of the Mercury-like 3:2 spin state, we need to keep the term containing $K_2(2\Omega_p-3n)$, as well as the leading term responsible for the bias, which is the $e^0$ part of the semidiurnal input:

%\begin{adjustwidth}{-\extralength}{-0cm}
%\centering %% If there is a figure in wide page, please release command \centering
\ba
\left(\frac{da}{dt}\right)^{\rm (pq,\s3:2)}\hspace{-6pt}=3\s a\s n\s \frac{M_m}{M}\left(\frac{R}{a}\right)^5 K_2(2\Omega_p -2n)\,+\,\frac{441}{8}\,a\,n\,e^2\,
 \frac{\;M_m}{M}\,\left(\frac{R}{a}\right)^5\,K_2(2{\Omega}_p-3n)
\;\,.
\label{}
\ea
%\end{adjustwidth}

Consider that this expression and all the subsequent analysis are valid for sufficiently small eccentricities because the series for $da/dt$ over the powers of $e$ is notorious for its slow convergence because of  large coefficients accompanying the higher-order terms in $\s e\s$.

The derivative of the above expression reads
\ba
\frac{d}{da}
\left(\frac{da}{dt}\right)^{\rm (pq,\s3:2)}_{
\,{\Omega\s}_{\!p}=3n/2
}= 3\s n\s\frac{M_m}{M}\left(\frac{R}{a}\right)^5\left[\,-\,\frac{11}{2}\s\left( K_2(2\Omega_p-2n)\s+\s\frac{147}{8}\s e^2\s K_2(2\Omega_p-3n)
\right)  \right.
\nonumber\\
\label{}\\
\nonumber
\left.
-\s3\s n\s\left(\frac{d\Omega_p}{dn}\s-\s 1  \right) \s\frac{dK_2(\omega)}{d\omega}\s\Bigg{|}_{\,\omega=2\Omega_p-2n}
-\, \frac{441}{8}\s n\s e^2\s\left(\frac{d\Omega_p}{dn}\s-\s \frac{3}{2}  \right)
\frac{dK_2(\omega)}{d\omega}\s\Bigg{|}_{\,\omega=2\Omega_p-3n}
\right]\;\,.
\ea

In the $2\Omega_p-3n=0$ resonance, the term $K_2(2\Omega_p-3n)$ vanishes, and we result with
\ba
\frac{d}{da}
\left(\frac{da}{dt}\right)^{\rm (pq,\s3:2)}_{
\,{\Omega\s}_{\!p}=3n/2
}= 3\s n\s\frac{M_m}{M}\left(\frac{R}{a}\right)^5\left[\,-\,\frac{11}{2}\s K_2(n)  \right. \qquad\qquad\qquad\qquad\qquad\qquad\qquad
\nonumber\\
\label{equa}\\
\nonumber
\left.
-\s3\s n\s\left(\frac{d\Omega_p}{dn}\s-\s 1  \right) \s\frac{dK_2(\omega)}{d\omega}\s\Bigg{|}_{\,\omega=n}
-\, \frac{441}{8}\s n\s e^2\s\left(\frac{d\Omega_p}{dn}\s-\s \frac{3}{2}  \right)
\frac{dK_2(\omega)}{d\omega}\s\Bigg{|}_{\,\omega=0}\;\;
\right]\;\,.
\ea

Since $K_2(n) >0$, the first term in square brackets is always negative and is working to make the equilibrium stable.
For realistic rheologies, $\;{dK_2(\omega)}/{d\omega}\s\Big{|}_{\,\omega=0}$ is positive, see e.g., Equation~(\ref{K2}). Therefore, for $d\Omega_p/dn > 3/2$, the third term also is negative and, therefore, {stabilising}.

The case of the second term is nontrivial. For semimolten viscous objects, the frequency $\omega=n$ may, in principle, reside to the left of the maximum of the $K_2(\omega)$ function given by {Equation}
%MDPI: We renamed ``Figure'' to ``Equation'', according to the contents of Appendix and the linked citation. Please check and confirm if this is correct or if Figures should be added to Appendix.
%  Authors: confirmed
~(\ref{K2}) {and depicted in Figure \ref{Fig1}}. In this case, the value of the derivative $\;{dK_2(\omega)}/{d\omega}\s\Big{|}_{\,\omega=n}$ is positive, and the second term in Equation~(\ref{equa}) is negative for $d\Omega_p/dn > 1\s$. Summing up, for $d\Omega_p/dn > 3/2\s$, all terms are negative and the 3:2 spin state is stable. For  $1 < d\Omega_p/dn < 3/2\s$, the result is likely to be the same, because the now-positive third term is unlikely to beat the first two (aside from $e^2$, it contains $\;{dK_2(\omega)}/{d\omega}\s\Big{|}_{\,\omega=0}$ which is not too large for semimolten bodies.)

For cold terrestrial bodies, the slope of $K_2(\omega)$ near $\omega=0$ is extremely steep. At the same time, for such bodies the value $\omega=n$  {resides} well to the right of the peak in Figure \ref{Fig1}, so the derivative $\;{dK_2(\omega)}/{d\omega}\s\Big{|}_{\,\omega=n}$ is negative, and the function $K_2(\omega)$ is slowly falling off as $\omega^{-\alpha}$, $\alpha=0.1 - 0.3$, for Andrade mantles, or as $\omega^{-1}$ for Maxwell mantles; see Equation~(\ref{555}) below. Under these circumstances, the last term in Equation~(\ref{equa}), {despite being proportional to $e^2$}, is likely to overpower the penultimate term. Given that the first term in square brackets is negative, we can say that, in this situation, the condition $d\Omega_p/dn > 3/2$ serves as a warranty of the 3:2 spin state being stable.

To sum up, we may expect that, in realistic settings, the condition $d\Omega_p/dn > 3/2$ ensures the stability of the 3:2 spin--orbit resonance.  Whether or not the resonant state is stable for $d\Omega_p/dn < 3/2$ needs to be explored on a case-by-case basis.

\section{Analytical approximation of the rate of spin-orbital evolution}

\begin{figure}[H]
 \vspace{2.5mm}
% \centering
 \includegraphics[angle=0,width=0.68\textwidth]{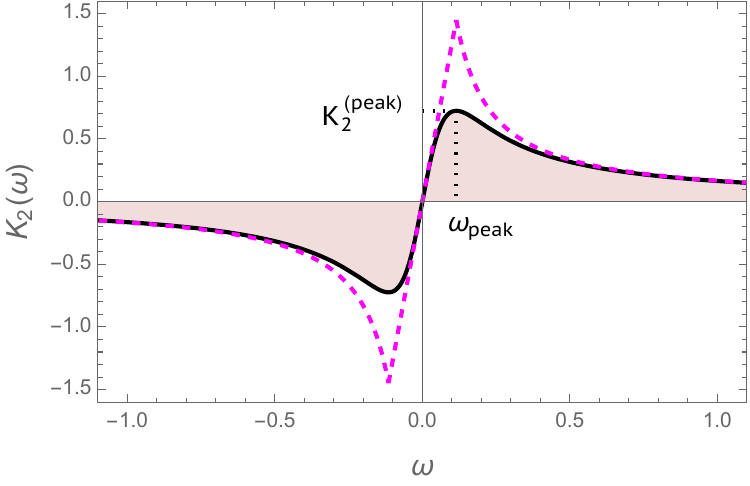}
 \caption{A typical form of a quadrupole quality function $ K_{\textstyle{_2}}(\omega)$, ~where $\,\omega\,$ is a shortened notation for the tidal Fourier mode
 $\,\omega_{\textstyle{_{2mpq}}}\,$. \,Specifically,  $\s\omega=\omega_{2200}\s$ is the principal, semidiurnal tidal mode
 that is shown with the solid black curve. The dashed magenta {curve} represents the piecewise function defined by Equations (\ref{444}) and (\ref{555}) and approximating the $K_2(\omega)$ function for the purpose of estimating the rate of orbital evolution outside the main resonance. Its peaks are twice the corresponding peaks of $K_2(\omega)$.
  \label{Fig1}}
 \end{figure}

The tidal quality function of a homogeneous spherical Maxwell body is given by Equation~(\ref{K2}). This function has the shape of a kink, as in Figure~\ref{Fig1}.

 It ensues from Formula~(\ref{K2}) that the maximum and minimum of the kink are attained~at
 %  \begin{linenomath*}
  \ba
{\omega_{\rm{peak}}}
 %_{\textstyle{_l}}
\,=\;\pm\;
 \;\frac{\tau_{\rm{_M}}^{-1}}{1\,+\,{\cal{A}}_2}\;
 %  \approx\;\pm\;\frac{1}{{\cal{B}}_l\,\eta}~~,
 \label{wh}
 \ea
 %  \end{linenomath*}
 and assume the values  \footnote{~The peak values given by Equation {(\ref{see})} depend, via ${\cal A}_2$, on the rigidity $\mu$, but bear no dependence on the viscosity.  At the same time, the inter-peak distance is inversely proportional to the viscosity, owing to the presence of $\tau_M^{-1}$ in   Equation~(\ref{wh}).   Hence, the lower the viscosity, the higher the peak frequency $|{\omega_{\rm{peak}}}|\s$. This is why an increase in the temperature results in an increase of the inter-peak spread in Figure~\ref{Fig1}, with no change in the peaks' height.}
  % \begin{linenomath*}
 \ba
 K_2^{\rm{(peak)}}\s=\;\pm\;\frac{3}{4}\;\frac{ {\cal A}_2 }{ 1 + {\cal A}_2 }\,\;.
 \label{see}
 \ea
  %  \end{linenomath*}

The quality function $K_2(\omega)$ is near-linear between the peaks:
 %  \begin{linenomath*}
 \ba
 \label{444}
 K_2(\omega)\;
 \;\simeq\;\frac{3}{2}\;\frac{{\cal A}_2}{1\,+\,{\cal A}_2}
 \;\frac{\omega}{|\s{\omega_{\rm{peak}}}\s|}~\,~,\quad\mbox{for}\qquad
 |\s\omega\s|\;<\;|\s{\omega_{\rm{peak}}}\s|
 \,\;,
 \ea
 %  \end{linenomath*}
and falls off as about $\omega^{-1}$ outside the said interval:
 %  \begin{linenomath*}
 \ba
  K_2(\omega)
 \;\simeq\;\frac{3}{2}\;\frac{{\cal A}_2}{1\,+\,{\cal A}_2}\;\frac{|\s{\omega_{\rm{peak}}}\s|}{\omega}~\,~,
 \qquad\mbox{for}\quad
 |\s\omega\s|\;>\;|\s{\omega_{\rm{peak}}}\s|\;\;,
 \label{do}
 \label{555}
 \ea
 %  \end{linenomath*}
where the factor $\,\frac{\textstyle 3}{\textstyle 2}\;\frac{\textstyle {\cal A}_2}{\textstyle 1\,+\,{\cal A}_2}\s$ is twice the actual maximum given by expression (\ref{see}).

For rigid and inviscid planets like the Earth, $\s |\s\omega_{\rm peak}\s |\s$ is close to zero, so the linear segment of the function between the peaks is steep, and capture into the main spin--orbit resonance happens relatively quickly once the system transcends the peak frequency. The subsequent  orbital evolution is then governed by the linear-in-frequency tidal reaction (\ref{444}) known as the Constant Time Lag (CTL) model. This regime has {often} been addressed in the literature, and we shall not dwell on it---especially because the resonance is very narrow for inviscid planets and is effectively a synchronous state with free librations that decay~rapidly.

{The protracted phase described by Equation~(\ref{555}) is of a greater interest for this study}, because this phase comprises the dynamics of approach to synchronism from a distant initial state. Combining Equation~(\ref{555}) with expression (\ref{term3}) we obtain, in the quadrupole approximation (and in the leading order of $e$),
\be
a^4\dot a (1-\Omega_p/n)=-\alpha,
\ee
where
\be
\alpha\s=\,\frac{9}{4} \frac{\;M_m}{M}\,R^5\,
%  \hat{K}
\frac{{\cal A}_2}{1\,+\,{\cal A}_2}
\,|\s\omega_{\rm peak}\s|
\;=\;
\frac{9}{4} \frac{\;M_m}{M}\,R^5\,
%  \hat{K}
\frac{{\cal A}_2\,\tau_M^{-1}}{(1\,+\,{\cal A}_2)^2}
\;\,.
\ee

Using the angular momentum conservation law in the form of Equation~(\ref{5}), we express $\Omega_p$ via $\sqrt{a}$ and arrive at the following differential equation:
\be
\frac{d}{dt}\left( \frac{X}{14}\,s^{14}-\frac{C}{13}\,s^{13}+\frac{\sqrt{G M_{\rm tot}}}{10}\,s^{10}\right)
= \,-\;\frac{\sqrt{G M_{\rm tot}}}{2}\; \alpha\;\,,
\ee
where we again use a shorthand, $s\equiv \sqrt{a}$. This equation {leads to the following solution}:
\be
\frac{X}{14}\,s^{14}-\frac{C}{13}\,s^{13}+\frac{\sqrt{G M_{\rm tot}}}{10}\,s^{10} = -\frac{\sqrt{G M_{\rm tot}}}{2}\; \alpha t + B\;\,,
\label{polynomial}
\ee
where $B$ is a constant of integration. By setting $t=t_0$, it is easily linked to the initial condition $s(t_0)$.

Computations with  {\it {Mathematica}} show that, for realistic {parameter} values, Equation~(\ref{polynomial}) has fourteen roots, of which only two happen to be real outside a close vicinity of the synchronous resonance, and only one (root number 14) turns out to be positive. This equation allows us to calculate the value of $a$ at any time in the past or future of the initial state $\{a(t_0), \Omega_p(t_0)\}$, as long as the planet--moon system remains outside the synchronous spin--orbit resonance, and insofar as the eccentricity is small.\,\footnote{~Recall that, in Equation~(\ref{term3}), we took into account only the leading input into the semimajor axis' rate---a simplification permissible at low eccentricities only.}
~It also enables us to calculate the time required for the system to evolve from an initial condition $\s s(t_0)=\sqrt{a(t_0)}\,$ to some $\s s(t)=\sqrt{a(t)}\,$:
\ba
t - t_0
&=& \frac{2}{\alpha}
\left(
\frac{X}{14\,\sqrt{G\, M_{\rm tot}}}\,\left[\s s(t_0)^{14}\s-\s s^{14}\s\right]
\right.
\nonumber\\
\label{difference}\\
\nonumber
&\;&-\;
\left.\frac{C}{13\,\sqrt{G\, M_{\rm tot}}}\,\left[\s s(t_0)^{13}\s-\s s^{13} \s\right]+\frac{1}{10}\,\left[\s s(t_0)^{10}\s-\s s^{10}\s\right]\s
\right)
\;\,.
\ea

For Pluto's mean radius and density, and with the rigidity assumed to be typical of ices, $\mu=4\times 10^9$ Pa, Equation~(\ref{A2}) gives ${\cal A}_2=28.02$, wherefrom $\,\frac{\textstyle 3}{\textstyle 2}\;\frac{\textstyle {\cal A}_2}{\textstyle 1\,+\,{\cal A}_2}\s
% \hat{K}
\approx\s 1.45\s$.
Assuming  $\s\omega_{\rm peak}=0.0005\,n\s$, we find from Equation~(\ref{polynomial}) that with the initial state $s(0)=5000$~m$^{1/2}$, Charon would have descended to $s=4998$~m$^{1/2}$ in the first 1 Myr. With the assumed parameters, {it would take the Pluto--Charon system 105 Myr to reach the current spin--orbit~equilibrium.}

\section{The impact of tidal perturbation due to the star}

A two-body planet--moon tidal system is a simplified model that neglects the influence of the star. In real star--planet--moon systems, this simplification is not always acceptable, primarily due to a secular torque acting on the planet from the tidal bulge raised on it by the star. When the planet's rotation is not synchronised with the moon's mean motion, the star-generated torque on the planet changes the slope of the tidal evolution track. However, if the planet becomes successfully synchronised by the moon, what is the net effect of the remaining secular torque from the star?

Figure~\ref{toy.fig} illustrates this dynamical consideration graphically. In the chosen axes, $\sqrt{a}$ and $\Omega_p$, given in arbitrary units, the locus of the synchronisation condition is represented by the red curve, while a possible evolutionary track due to the tides from the moon alone is shown by the solid blue line. The black disk marks the position of the long-term stable tidal equilibrium when the planet rotates synchronously with the orbital motion of the moon. The system can reach this equilibrium from either direction along the solid line.
{Unless the planet is synchronised by the star, the planet's orbital motion is normally expected to be slower than its spin. Therefore, the star} gives rise to a leading tidal bulge on the planet, unless the planet's rotation is retrograde. This effects acts to decelerate the planet's rotation. Assume, within a toy dynamical model, that this external action is an impulse, which instantaneously reduces $\Omega_p$, as shown with a downward arrow originating from the black disk. Since the actual rate of rotation is now slower than the synchronisation value, a tidal torque from the moon emerges, which moves the system along the new linear track (the first dashed line below) up and to the left, until the new point of intersection is attained and the synchronisation equilibrium is restored. A new braking impulse from the star starts another cycle of evolution, as shown with the corresponding pair of arrows. This process continues until the newly acquired tidal track intersects the cubical curve in a single tangent point (shown with a red circle). As we previously noted, this equilibrium is unstable, and the system will continue to shrink forever.

To transit from this toy model to actual physical behaviour, let us make the braking impulses from the star continuous and infinitesimal. It is obvious that, as long as the tidal action from the moon is stronger than that from the star, the net effect will be a secular drift of the system from the initial point (black disk) along the cubical curve towards the critical tangent point (red circle). The braking tidal torque on the planet from the star is transformed by this mechanical configuration into a braking moment on the moon and faster rotation of the planet via the stronger tidal response from the moon. If the tangent point is ever reached, the planet becomes suddenly liberated from the synchronisation lock, and the moon will commence its final descent. Thus, in the very long run, a synchronisation state is transient, with the system drifting along the cubical curve track towards the single point of unstable equilibrium, and then along the linear track towards the final downward spiral. If such previously synchronous moons are ever found, the current $(\!\sqrt{a}\s,\,\Omega_p)$ state can be used to compute the slope $X$ in Equation~(\ref{X}) and, therefore, to estimate the moon's~mass.

\begin{figure}[H]
    \includegraphics[width=0.75\textwidth]{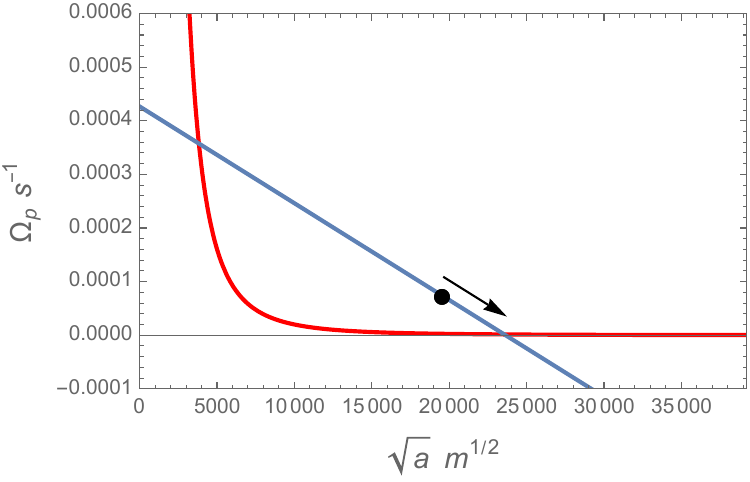}
    \caption{{Tidal evolution} 
 of the Earth--Moon system in the $(\sqrt{a}\s,\,\Omega_p)$ axes. In accordance with Equations~(\ref{Omega}) and (\ref{OmegaS}), the straight line is an evolutionary track, while the cubical curve comprises the available synchronous states. The direction of evolution along the straight line is shown with an inclined arrow. The filled circle denotes the current state of the system, taken as an initial condition. The intersection of the straight line with the cubical curve, to the right of the initial condition, is a point of synchronous equilibrium.}
    \label{moon.fig}
\end{figure}

{The impact of the tides raised by the star in the general non-equilibrium configuration of an expanding orbit and prograde motion (the Sun--Earth--Moon case depicted in Figure~\ref{moon.fig}) is also a systematic perturbation of the planet's spin, which is aligned with the braking moment from the moon. The trajectory will be slightly bent downward in this Figure. The main effect is that the point of stable equilibrium will be reached sooner and at a slightly shorter separation.}

\section{Examples}

Condition (\ref{cond.eq}) can be used to estimate the chances of non-synchronous planet--moon systems to synchronise, and to map the initial parameter space of currently synchronised systems. The latter is relevant for the still unclear origins of planet--moon systems. To illustrate the utility of condition (\ref{cond.eq}), we apply it to several well-known systems.

\subsection{The Earth--Moon system}

Mutual synchronism as an end state of a tidally interacting Earth--Moon system was proposed on qualitative grounds by Immanuel~\citet{kant1754} in an article that appeared on {8 and 15 June 1754}
%MDPI: We revised date format according to our journal requirements. Please confirm.
%Authors: OK
, across two issues of a K\"{o}nigsberg newspaper. This was Kant's first journal publication and his second publication overall. Republished almost a century and a half later in Kant's {``$\rm Gesammelte~Schriften$''~\citep{kant1900}, this work remains virtually unknown among~astronomers.

The Earth is presently rotating with a high angular velocity of $\Omega_p=7.27\times 10^{-5}$~s$^{-1}$, and undoubtedly had a faster spin in the past, when the Moon was closer to it. This is the most important example of the spin-down/expansion scenario. Can the Earth be synchronised by the Moon in principle, or will the Moon drift into infinity with the remaining finite rotation of the planet? To answer this question, we use the current Earth rotation rate $\Omega_p(t_0)$ and the current semimajor axis $a(t_0)$ of the Moon as the initial values, and compute both sides of inequality (\ref{cond.eq}). For the realistic value of $\xi=1/3\s$, Formulae~(\ref{X}), (\ref{C}), and (\ref{Csyn}) render $X=1.80\times 10^{-8}$ m$^{-{1}/{2}}$ s$^{-1}\s$, $C=4.27\times 10^{-4}$~s$^{-1}$, and $\,C_{\rm syn}=1.83\times 10^{-4}$~s$^{-1}\s$. Synchronisation condition (\ref{cond.eq}) is then fulfilled by a large margin.

This estimate excludes the contribution from the solar tides on Earth, which are weaker than the lunar tides. In the considered setting, the tidal action from the star is directed along the action from the moon because it contributes to the spin-down of the planet. {This situation may be different for close exoplanets with massive moons, where the tidal torques from the star and the satellite can be counter-directed for slowly rotating planets.} If the condition is barely fulfilled for a given configuration, the contribution from the star-raised torque should be taken into account.

Figure~\ref{moon.fig} depicts the evolution of the Earth--Moon system toward the point of synchronisation in the $\s\Omega_p\s$ versus $\s\sqrt{a}\s$ axes. The red cubical curve is the locus of all synchronous states, the inclined straight line is the trajectory of the system, and the filled circle is the current state. The system evolves along the straight line toward the second intersection point, as indicated by the arrow. If this point is reached, the system practically stops evolving, because the time-averaged total tidal torque on the synchronously rotating Earth vanishes and the solar tidal torque is unable to drive the Earth from this equilibrium (for details, see Section 9.1.1 in {reference} %MDPI: Newly added, please confirm. The same below.
%  OK
~\cite{2023A&A...672A..78M}). We note that the left-side intersection of the curves also provides a synchronisation equilibrium (an unstable one). If the Earth evolved in the past from much faster $\Omega_p$ values, could it have crossed that unstable equilibrium? The answer to this question is negative, because the synchronous radius value $a_s$ for that state is only $\approx2.2\;R_{\rm Earth}$, which is within the Roche radius. The Moon was never that close to the Earth. Furthermore, the section of the tidal track to the left of the first point of equilibrium is below the synchronisation rate $\Omega_s\s$, and the system would evolve away from this state.

The value of $a$ in the right-side crossing point, a stable synchronous state, can be found by solving Equation~(\ref{quartic}) and is about $5.56\times 10^8$ m $=\s87.1\s R_{\rm Earth}$. This value falls short of Earth's reduced Hill radius, which is about $111\s R_{\rm Earth}$. In reality, this synchronous value will never be attained, because the required time exceeds the life expectancy of a stellar system.\,\footnote{~The current rate of Earth's rotation period is $\s- 1.7$ ms/century. At such a pace, it will take Earth about $100$ Bln years to reduce its spin to the present mean motion of the Moon. By then, the Moon will be residing much further than now, and its mean motion will be much slower; so synchronisation will require even more time.}
The Moon's spin synchronisation time is, in any realistic scenario, orders of magnitude shorter than Earth's, ranging from a few million to tens of millions of years. The general reason for this can be understood from, e.g., Formulae~(D1) and (D2) in the appendix of {reference}~\cite{Mars}.
For a detailed discussion and numbers, see~\cite{2013MNRAS.434L..21M} and the references therein.

\subsection{Neptune--Triton}

Triton is the only large moon in the solar system with a retrograde orbital motion with respect to the spin direction of the host planet. The origin of this unusual satellite is still debated, with the leading hypothesis suggesting a dynamical capture via a gravitational encounter with a binary planet on a heliocentric orbit~\citep{2006Natur.441..192A}. The analytics of Section~\ref{con.sec} are fully applicable here, with the necessary modification that the initial spin rate $\Omega_p(t_0)$ is negative for the positive definite orbital motion $n$.

Employing Neptune's present-day rotation rate $\Omega_p(t_0)=-1.08\times 10^{-4}$ s$^{-1}$ and Triton's present semimajor axis $a(t_0) = 3.55 \times 10^8 $ m as the initial conditions, and using for Neptune the approximation $\xi=1/3\s$, we obtain from Formulae~(\ref{X}), (\ref{C}), and (\ref{Csyn}) $X=8.5\times 10^{-11}$ m$^{-{1}/{2}}$ s$^{-1}\s$, $\,C=-1.07\times 10^{-4}$ s$^{-1}\s$, and $C_{\rm syn}=4.7\times 10^{-6}$ s$^{-1}$. The quantity $C$ is negative for this system due to the low angular momentum transfer rate $X$ and the fast present spin of Neptune in the direction counter to Triton's apparent mean motion. Triton is destined to perish in Neptune's Roche zone in the very long run, while Neptune's rotation rate will barely change. We note that the actual transfer of the orbital momentum of Triton to Neptune's spin will be even more limited because of the substantial ($\sim23${\textdegree}%MDPI: We changed superscript letter ``o'' with the degree symbol in the paper. Please confirm.
) obliquity of the orbit on the equator. The time required for Triton to spiral down is longer than the anticipated lifetime of the solar system, so this scenario is somewhat scholastic. It may nonetheless be relevant for our understanding of the rarity of retrograde moons within the known planetary systems.

\subsection{Venus--Neith}

\citet{mccord1968},~\citet{1970Sci...170.1196S},~\citet{burns1973},~\citet{counselman1973},~\citet{ward1973},~\citet{malcuit1995}, and~\citet{2023Univ...10...15M} hypothesised in their works that both the unusually slow rotation rate and the retrograde rotation direction of the near-twin planet Venus can be explained by its tidal interaction, in the past, with a putative Moon-like satellite called Neith. It was specifically demonstrated by~\citet{2023Univ...10...15M} that the satellite could have been stochastically captured by Venus from an independent orbit around the Sun into a retrograde orbit around Venus through a chaotic layer at the Hill radius. The validity of this process has been demonstrated in the literature~\citep{2003Natur.423..264A}. While a direct capture in the hierarchical Sun--Venus--Neith setup appears to have a low probability on a single encounter, a triple-encounter configuration can enhance the chances of a long-term recombination~\citep{2006Natur.441..192A, 1971SvA....15..411A}. The probability of capture is also increased by the presence of a primordial circumplanetary nebula~\citep{1979Icar...37..587P}, which can be efficient in bringing down the moon from the initial separation of $\sim100\s R_V\s$, where the external capture takes place, to a much closer critical distance of $\sim8\s R_V\s$, where the tidal dissipation forces begin to dominate. Particles in the primordial torus are expected to orbit Venus in the prograde direction (with respect to the general solar system spin and Venus' initial rotation). The captured moon is moving in the opposite direction, increasing the rate of gas drag dissipation and swiftly sweeping up the circumplanetary material.

After having approached Venus sufficiently closely, Neith synchronised its spin with the orbital motion, but the planet kept rotating in the direction opposite to Neith's orbit. The tidal bulge on Venus raised by Neith moved across the surface at a considerable angular rate, combining the planet's sidereal spin and the orbital motion. The moon's orbit decayed, while the planet's spin slowed down and eventually reversed. This process terminated when Neith reached the Roche radius (approximately, $2R_V$), where it was disrupted by the internal tidal tension. Within this scenario, the moon's mass should not be too large, because a massive retrograde satellite would probably synchronise Venus before reaching the Roche radius, and we would still enjoy a spectacular close moon rapidly revolving around Venus. Likewise, the initial prograde spin rate of Venus should be above a certain limit to preclude this outcome. Using the condition inequality derived in this study, we can quantify the range of acceptable parameters.

In this case, the evolutionary track of the planet--moon system is pinned to the current configuration, which is assumed to be close to the end-point at the time of Neith's destruction. With that, we set $\Omega_p(t_0)=2.99\times 10^{-7}$ s$^{-1}$ and $\sqrt{a(t_0)}=4261$~m$^{1/2}$. Note that the sidereal spin rate of the planet is positive here, because the orbital mean motion of the retrograde moon is positive-definite in the derivation of the synchronisation condition. Figure~\ref{neith.fig} shows the corresponding condition limit and momentum exchange trajectories for this scenario. The red open circle indicates the end-point of the Venus--Neith system at the time of Neith's destruction. The two straight dashed lines intersecting at this point are the tidal evolution tracks for assumed Neith's masses $1\,M_{\rm Moon}$ (shallower) and $5\,M_{\rm Moon}$ (steeper). These tracks miss the synchronisation curve $\Omega_s\s$, depicted with a red cubical hyperbola, by a large margin. Irrespective of the satellite's mass, the synchronisation of Venus was impossible with the assumed end-point conditions. The tidal tracks allow us to put useful constraints on the initial rate of Venus' rotation in the past. Indeed, assuming that the initial separation was $8\,R_V$ at the time of circumplanetary nebula dispersal, we can deduce from this plot that the sidereal period of rotation (in the prograde sense) was 29 hr for $M_{\rm Neith}=1\,M_{\rm Moon}$ and 6 hr for $M_{\rm Neith}=5\,M_{\rm Moon}$. The latter estimate appears to be more plausible (also increasing the speed of tidal decay) in comparison with the present-day rates of rotation of the solar planets. We can also evaluate the requirements for the case of surviving retrograde moon, which would be able to synchronise Venus before reaching the Roche radius. The dotted straight line in Figure~\ref{neith.fig} shows the marginal trajectory for $M_{\rm Neith}=1\,M_{\rm Moon}$ touching the red hyperbola in one point. We find that this low-mass synchronisation condition corresponds to a rate $\Omega_p\sim 5\times 10^{-5}$ s$^{-1}$ (the positive sign indicating retrograde rotation), or a rotation period of 35 hr. In a more distant past, a very slow prograde rotation of Venus would be needed to realise the synchronisation option.

\begin{figure}[H]
    \includegraphics[width=0.75\textwidth]{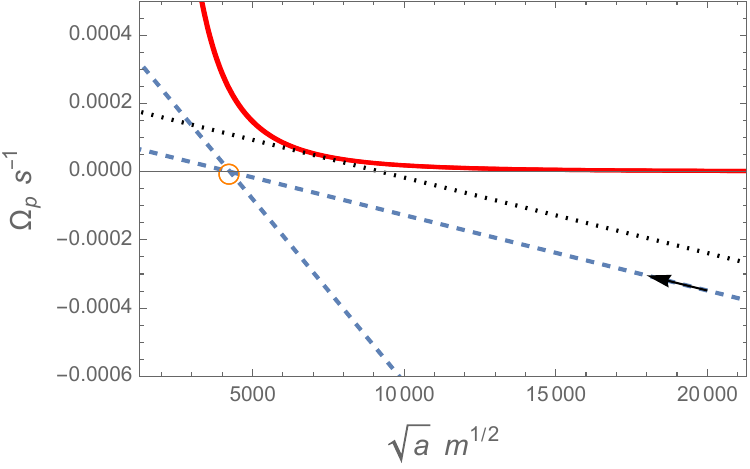}
    \caption{{Tidal evolution} of the hypothetical Venus--Neith system in the $\left(\sqrt{a}\s,\,\Omega_p\right)$ axes. The end-state of the system at the time of Neith's destruction is marked with an {orange} open circle. The arrow shows the direction of evolution along the tidal angular momentum exchange track for an assumed $1\,M_{\rm Moon}$ mass of the satellite. The steeper dashed line shows the track corresponding to a more massive satellite at $5\,M_{\rm Moon}$. The dotted black line shows the computed track for $M_{\rm Neith}=1\,M_{\rm Moon}$ option that marginally fulfills the synchronisation condition (\ref{cond.eq}).}
    \label{neith.fig}
\end{figure}

\subsection{Mars, Phobos, and Deimos}

The two enigmatic moons of Mars, Phobos and Deimos, are certainly too small to perceptibly change the rotational momentum of Mars. It may seem that the answer to the question of possible synchronisation is predetermined. The contribution from both moons to the total angular momentum of the system is negligible due to their small mass, and the values of $C$ for these satellites, as given by Equation~(\ref{C}), are dominated by the high angular momentum of the planet. Indeed, the insertion of Mars' present spin rate $\Omega_p(t_0)=7.095\times 10^{-5}$ s$^{-1}$, along with the present semimajor axis value $a(t_0)$ for each moon, demonstrates that, for both moons, the condition of synchronisation (\ref{cond.eq}) is fulfilled, so either moon can, {\it{formally}}, be in a stable synchronism with Mars' rotation.

To clarify this point, consider the functions (\ref{Omega}) and (\ref{OmegaS}) depicted in Figure~\ref{phobos.fig}. For both Phobos and Deimos, the tidal track is shown with a dashed straight line. For each moon, the line is nearly horizontal. Its weak incline is barely noticeable in the figure, because the tidal evolution only entails minuscule changes in the planet's spin rate. Nevertheless, owing to this weak incline, the track intersects the locus of synchronous states (red hyperbola) not in one, but in two points. One intersection point, that of a stable synchronism, is residing away to the right, at a distance much exceeding the Martian Hill radius---for which reason the configuration cannot be implemented in practice. However, within the range of practical interest is located the second intersection point, that of unstable synchronicity. Depicted in the figure, it corresponds to Mars' present synchronisation radius, $r_{\rm sync}$ = 20,415 km. Both moons are now moving along the tidal track in the opposite directions and away from the synchronisation point, as shown with black arrows. Phobos is relatively quickly falling onto the planet, and Deimos is slowly receding.

\begin{figure}[H]
    \includegraphics[width=0.6\textwidth]{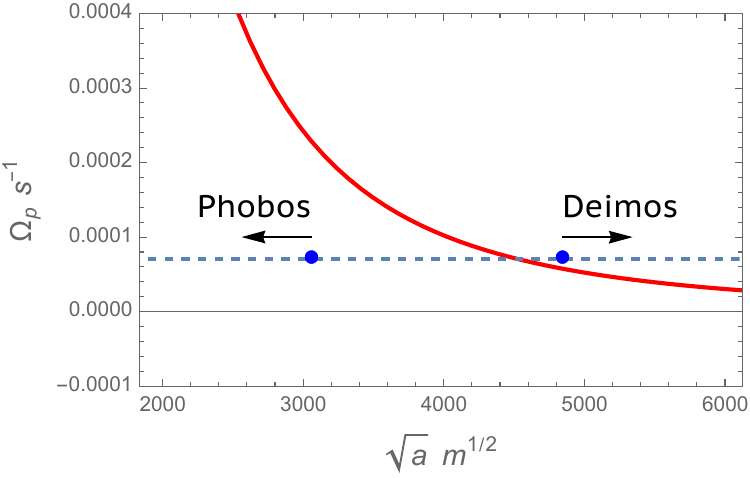}
    \caption{Tidal evolution of the Mars--Phobos--Deimos system in the $(\sqrt{a}\s,\,\Omega_p)$ axes. The current states of Phobos and Deimos are indicated with blue circles, and the directions of their evolution along the common linear track are shown with black arrows. The synchronisation condition (\ref{cond.eq}) is achieved for either satellite at the point of intersection of the cubical hyperbola with the tidal track, where the corresponding semimajor axis equals the synchronisation radius of the system. That equilibrium is intrinsically unstable.}
    \label{phobos.fig}
\end{figure}

Some authors suggest that Phobos and Deimos could have been either formed {in situ} %MDPI: We removed italics form latin expressions according to journal requirements, please confirm.
%Authors: OK
 or captured from outside the Hill sphere~\citep{2024}. The capture theory faces challenges in explaining the current state of the system, particularly the low eccentricity of Deimos~\citep{1983Icar...55..133S}. Within the {in situ} formation theory, the material for new moons in the outer regions of the swept-up debris disk could have been delivered by the last giant impact of Mars~\citep{2016NatGe...9..581R}, and later shaped into the current configuration by passing planetesimals of larger mass~\citep{2018MNRAS.475.2452H}. This scenario presents its own set of problems from the dynamical standpoint. What happened to the other multiple moonlets formed in the hypothetical debris disk? How could Phobos and Deimos become so much separated by random impulsive interactions with bodies moving on hyperbolic orbits, which endow initially circular Phobos's and Deimos's orbits with higher eccentricity?

A more elaborate hypothesis is that the moons of Mars originated from a significantly more massive satellite, a common progenitor crushed by an external projectile~\citep{1979Icar...37..587P, 2021NatAs...5..539B}. Given that Phobos and Deimos are moving away from a point of unstable equilibrium, could Mars have been synchronised by the progenitor body? This body would be at $r_{\rm sync}$ at the time of impact. A simple calculation reveals that this is not possible. The total orbital energy of Deimos today is $1.4\times 10^{21}$ J, which is much greater than the total energy of Phobos ($2.4\times 10^{17}$ J). Subsequently, the energy required to lift Deimos from $r_{\rm sync}$ to its current position is $2.1\times 10^{20}$~J, while the energy released by shifting Phobos inward is smaller by a few orders of magnitude. Thus, the progenitor body should have resided well above $r_{\rm sync}$ at the time of impact, and quite close to the current orbit of Deimos.

\subsection{Pluto and Charon}
\label{p-c.sec}

%  \textls[-25]
{Pluto and Charon represent a remarkable case of successful mutual synchronisation}~\citep{Buie1997, Buie2010}. Although the tidal despinning of Charon is faster than that of Pluto, it does not rule out the possibility of Charon having stayed in a stable higher spin--orbit resonance in the past, when the orbital eccentricity was much higher than presently~\citep{Amirs}.

In the framework of this study, the initial conditions $\Omega_p(t_0)$ and $\sqrt{a(t_0)}$ should be pinned to the current state, which is the point of intersection of the corresponding functions (\ref{5}) and (\ref{6}). This configuration is illustrated in Figure~\ref{pluto.fig}.

Since the mass of Charon is not negligible at 12\% of the mass of Pluto, one may question the validity of our simplifying assumption that the tidal evolution track is defined by the momentum transfer between the orbit and the larger body only. In Figure~\ref{pluto.fig}, we show two tracks, which correspond to two one-sided approximations. The solid blue line is a theoretical evolution of the system dynamics, with the angular momentum of Charon ignored. The dashed straight line is the track when the orbit evolves due to tidal dissipation of Charon's rotational energy only. This line is almost vertical, because Charon alone can hardly change the orbit. In a distant past, both bodies were rotating out of synchronism, but Charon got synchronised (or temporarily captured into a higher resonance) much faster than Pluto. After that, the orbit continued to change---and a tiny portion of angular momentum continued to be transferred to or from Charon---until Pluto became synchronised also.

In the general case, when both bodies are not rotationally synchronous, the angular momentum conservation can be cast in the form
\be
\dot n = \frac{3 \xi}{a^2 q_1 q_2}\, \left(q_1 R_1^2 \dot\Omega_1+q_2 R_2^2 \dot\Omega_2\right),
\ee
where we used index 1 for Pluto and 2 for Charon and used $q_i=M_i/(M_1+M_2)$ to denote the mass fractions. We also assumed, for simplicity, that both bodies have the same value $\xi$. The constant factors $q_i\,R_i^2$ for Pluto and Charon are in proportion \,31.5\,:\,1\,. Despite this, the contribution of Charon to the tidal evolution of orbit is considerable because Charon's spin-down rate much exceeds Pluto's spin-down rate (this can be easily understood by comparing Formulae~(D1) and (D2) from~\cite{Mars}). When Charon becomes stably synchronous, $\dot\Omega_2$ equalises with $\dot n$  and becomes much smaller. At this point, Charon practically stops contributing to the orbital evolution of the system. Therefore, the inclined track in Figure~\ref{pluto.fig} should faithfully represent the evolution path of the system in the vicinity of synchronisation equilibrium. But which direction had been taken by the system before the permanent capture produced the current state?

The arrow in Figure~\ref{pluto.fig} illustrates the scenario wherein the two bodies were initially closer and rotated at higher rates. Commonly accepted in the literature~\citep{2014Icar..233..242C, 2020A&A...644A..94C}, this hypothesis is consistent with a giant impact proposed earlier as the origin of Charon~\citep{1989ApJ...344L..41M}. At first glance, this seems to be a foregone conclusion, because an alternative would entail a negative (i.e., retrograde with respect to the orbital spin) initial rotation of Pluto. We recall, however, that the inclination of Charon's orbit is 122\textdegree~to the plane of the ecliptic, i.e., both Charon's orbit and Pluto's present rotation are retrograde with respect to the spin of the solar system. The commonly accepted model of expanding orbit then inevitably leads to a strong conclusion that Pluto had initially been rotating in the retrograde sense at a considerable rate. This consideration casts a shadow of doubt on the giant impact or binary fission hypotheses of Charon's origin. The alternative model including the external capture of Charon by an initially prograde Pluto at a large angle with respect to the ecliptic is therefore viable.\,\footnote{~The present equatorial plane of Pluto must be close to the initial orbit of Charon  because the orbital angular momentum after the capture was much larger than the initial spin orbital momentum of Pluto.} The perturbative torque from the Sun can vary the obliquity of Pluto within $\sim20$\textdegree~only~\citep{1983Icar...55..231D}, so the planet may owe its current retrograde spin to the tidal interaction with an externally captured Charon. A detailed description of this process, with numerics and plots, will be provided elsewhere.

\begin{figure}[H]
    \includegraphics[width=0.75\textwidth]{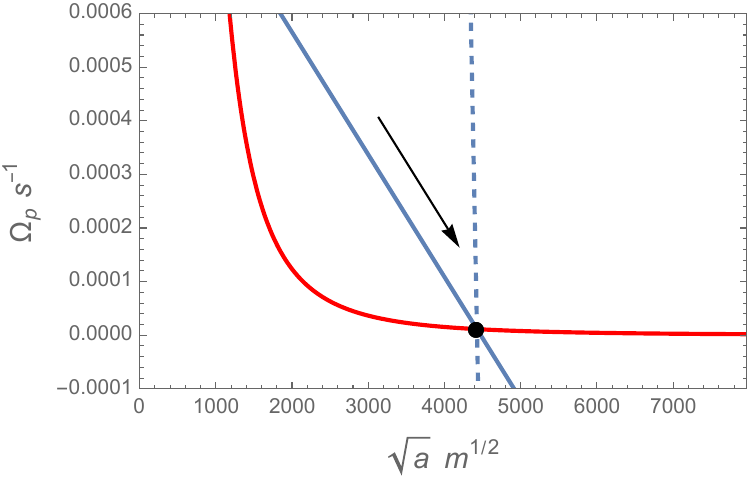}
    \caption{{Tidal evolution} %MDPI: Please add the explanation for red solid line in the figure.
 of the Pluto--Charon system in the $(\sqrt{a}\s,\,\Omega_p)$ axes. {The red hyperbola represents the states of spin--orbit equilibria.} The current state is indicated with a filled black circle, and the linear track of tidal evolution is shown with a blue solid line. The dashed line shows a hypothetical track when tidal dissipation takes place only in Charon. The arrow shows the commonly assumed direction of pre-equilibrium evolution, but the opposite evolution from higher orbits is physical and possible.}
    \label{pluto.fig}
\end{figure}

\section{Conclusions and questions}

In neglect of the angular momentum of the moon, and for small eccentricity and obliquities, we have derived a simple constraint on the physical parameters of a planet--moon system and the initial planet's spin rate and orbital separation, which ensures a possibility of planet's synchronisation with the tidally evolving orbit. This condition is universal, and bears no dependence upon the planet's internal structure or tidal dissipation model. It is applicable to dwindling systems, as well as tidally expanding orbits and cases of initially retrograde motion. The utility of this condition has been illustrated through examples of several well-known planet--moon pairs in the solar system.

We derive a constraint such that, when satisfied, the system can become completely synchronous in principle, although not always. When the condition is met, there are two possible values for the synchronous radius---those corresponding to a low-energy and a high-energy states of rotational equilibrium. The former is intrinsically unstable, while the latter is permanently stable. This is seen from the direction of angular momentum exchange in the axes $\left(\sqrt a\s,\,  \Omega_p\right)$ in Figures~\ref{toy.fig} and~\ref{moon.fig}. If an evolutionary track, which is a straight line with a negative slope, does not intersect the cubical curve representing synchronous states, the system always moves to the left and up; that is to say, the moon spirals down onto the planet. The same is true for the marginal case when the track is tangent to the hyperbola in one point, making this equilibrium intrinsically unstable. When the track intersects the cubical curve in two points, the segment between these points has the tidal torque directed toward a lower spin rate and a higher $a$. Therefore, for the stable high-energy synchronisation state, the tidal action provides a restoring torque whenever an external perturbation, independent of its direction, drives the system out of this state. The Earth--Moon system is heading toward its stable point of synchronism. The Mars--Phobos--Deimos system is diverging away from the low-$a$, unstable point of equilibrium, which suggests a super-synchronous configuration if they split from a single progenitor body or from a horseshoe co-orbital binary.

The Pluto--Charon pair is probably permanently captured into a stable, high-energy state of synchronism. Our analysis raises the question whether this pair arrived at this current state from an initially narrower configuration (as is commonly assumed) or from a wider orbit. The latter option would not require Pluto to have an initial retrograde spin with respect to the solar system angular momentum. It would be more conformal to the proposed external capture of Charon from an independent heliocentric orbit, rather than the giant impact scenario. Pluto and Charon would represent the case of a successful synchronisation of a small planet by an externally captured moon.

The hypothetical retrograde moon of Venus represents a case when our derived condition is not satisfied, and the moon inevitably spirals down onto the planet. Our specific calculations reveal that it is not the mass of Neith that was the critical factor for this scenario, but rather the initial spin rate of Venus in the prograde sense. A slow prograde spin rate with a period of $\sim35$~h would trigger the synchronisation condition for a retrograde moon of moderate mass, which would still be orbiting Venus today.

{The method of derivation of the local criterion provided in Section~\ref{criterion} can be generalised for an analysis in which other effects are included: a finite eccentricity value, higher-than-quadrupole terms, an input from the planet-caused tides in the satellite, etc.
For example, this analysis can be applied in the case of moons  which are not only tidally synchronised with the host planet, but are themselves in orbital resonance (like the Galilean moons). To explore the stability of the synchronous state of such a moon, one should also include into expression (\ref{term}) for $da/dt$ the leading terms emerging due to the interactions between these moons. An example of application for the case of 3:2 spin--orbit resonance is provided in Section~\ref{32.sec}}.

Our main results may find interesting applications in the studies of exoplanets. Owing to the fast-paced technical progress in space and ground-based observations, astronomers may be on the brink of major discoveries concerning rotational dynamics of exoplanets. If asynchronous exoplanets are detected, the option of tidal synchronisation by an exomoon should be considered, given the spectacular examples in our solar system. {The currently identified candidate exoplanet--exomoon system include Kepler 1708b-i~\citep{2022NatAs...6..367K} and Kepler 1625b-i~\citep{2018AJ....155...36T}, although the latter case has been disputed~\citep{2019ApJ...877L..15K}. In both cases, the suggested satellites of Uranus or Neptune mass orbit Jupiter-sized planets at a significant distance from their host planets. This configuration is closer to the Pluto--Charon case. Indirect evidence of a volcanic or evaporating moon was found in phase-shifted sodium lines in the spectrum of the WASP-49 A system~\citep{2024ApJ...973L..53O}, which would represent, if confirmed by follow-up observations, a rather improbable case of a rocky moon jammed in the narrow survival niche of the hot Jipiter WASP-49 Ab~\citep{2025A&A...694L...8S}.}

\clearpage

  \appendix
{\Large\bf APPENDIX}
\section[Derivation of expression (\ref{insynchro})]{}
\label{Appendix A}

\subsection[Semimajor axis' rate in spin-orbit synchronism]{Semimajor axis' rate in spin-orbit synchronism}

The semimajor axis' quadrupole tidal rate is \citep[Eqn 143]{Boue}:
 \ba
 \nonumber
 &\,&\left(\frac{da}{dt}\right)_{l=2}=
 ~\\
 \nonumber\\
 \nonumber
  &-&3\,a\,n\left(1\,-\,5\,e^2\,+\,\frac{63}{8}\,e^4\right)
 \left[\s\frac{\;M_m}{M}\,\left(\frac{R}{a}\right)^5\,K_2(2n-2{\Omega}_p)
 \right.
 ~\\
 \nonumber\\
 &~& \qquad\qquad\qquad\qquad\s\,\s\quad +\s\left. \frac{M}{M_m}\,\left(\frac{R_m}{a}\right)^5\,{K_2}_m(2n-2{\Omega}_m)
 \s\right]
 \nonumber
 ~\\
 \nonumber\\
 \nonumber
 &-&\frac{9}{4}ane^2\left(1+\frac{9}{4}e^2\right)
 \left[\frac{\;M_m}{M}\left(\frac{R}{a}\right)^5 K_2(n)
+
 \frac{M}{M_m}\left(\frac{R_m}{a}\right)^5{K_2}_m(n)\right]
 \nonumber
 ~\\
 \nonumber\\
 \nonumber
 &-&\frac{3}{8}\,a\,n\,e^2\,\left(1\,-\,\frac{1}{4}\,e^2\right)
 \left[\s\frac{\;M_m}{M}\,\left(\frac{R}{a}\right)^5\,K_2(n-2{\Omega}_p) \right.
 ~\\
 \nonumber\\
 &~& \qquad\qquad\qquad\qquad\s   + \left.
 \frac{M}{M_m}\,\left(\frac{R_m}{a}\right)^5\,{K_2}_m(n-2{\Omega}_m )\s\right]
 \nonumber\\
 \nonumber\\
 \nonumber
 &-&\frac{441}{8}\,a\,n\,e^2\,\left(1\,-\,\frac{123}{28}\,e^2\right)
 \left[\s\frac{\;M_m}{M}\,\left(\frac{R}{a}\right)^5\,K_2(3n-2{\Omega}_p) \right.
 ~\\
 \nonumber\\
 \nonumber
 &\s& \qquad\qquad\qquad\qquad\qquad   + \left.
 \frac{M}{M_m}\,\left(\frac{R_m}{a}\right)^5\,{K_2}_m(3n-2{\Omega}_m)\s\right]
 ~\\
 \nonumber\\
 \nonumber
 &-&\frac{867}{2}\,a\,n\,e^4\,
 \left[\s\frac{\;M_m}{M}\,\left(\frac{R}{a}\right)^5\,K_2(4n-2{\Omega}_p) \right.
 ~\\
 \nonumber\\
 \nonumber
 &\s& \qquad\quad\quad   + \left.
 \frac{M}{M_m}\,\left(\frac{R_m}{a}\right)^5\,{K_2}_m(4n-2{\Omega}_m)\s
 \right]
 \nonumber\\
 \nonumber\\
 \nonumber
 &-&\frac{81}{8}\,a\,n\,e^4\,
 \left[\s\frac{\;M_m}{M}\,\left(\frac{R}{a}\right)^5\,K_2(2n)
+
 \frac{M}{M_m}\,\left(\frac{R_m}{a}\right)^5\,{K_2}_m(2n)\s\right]
 \nonumber\\
 \nonumber\\
 &\s& \qquad\qquad\qquad\qquad   + O(i^{\,2})\,+\,O(i_m^{\,2})\,+\,O(e^{\,6})\;\,,\;\quad
 \label{ae1.eq}
 \ea
 where $a$, $n$, $e$, and $i$, $i_m$ are the semimajor axis, mean motion, eccentricity, and the partners' obliquities on the orbital plane.
 The notations $M$, $R$, $K_2$, and $\Omega_p$ are the mass, radius, quality function, and the sidereal rotation rate of the planet, while $M_m$, $R_m$, ${K_2}_m$, and $\Omega_m$ are their counterparts for the moon.

\subsection[Stability of synchronism, in neglect of the eccentricity and the tides in the secondary]{Stability of synchronism, in neglect of the eccentricity and the tides in the secondary}

In the above expression, single out the quadrupole semidiurnal term due to the tides in the planet:
\bs
\ba
\left(\frac{da}{dt}\right)^{\rm (pqs)} &=& -\;3\,a\,n
\frac{\;M_m}{M}\,\left(\frac{R}{a}\right)^5\,K_2(2n-2{\Omega}_p)
\label{term1}\\
&=& -\,3\frac{\,M_m}{M}\s R^5\s(GM)^{-4/3}\s n^{11/3}\s K_2(2n-2{\Omega}_p)\;,\;
\label{term2}
\ea
\label{term}
\es
``$\s(pqs)\s$'' meaning: {\it planet, quadrupole, semidiurnal}.

Thus ignoring the tides in the moon and neglecting higher-order (in the powers of $e$) contributions from the planet, we explore the influence of term (\ref{term}) on the stability of spin-orbit synchronism. To that end, we calculate the full derivative of expression (\ref{term2}) with respect to $a$ near synchronism:
\ba
\nonumber
\frac{d}{da}
\left(\frac{da}{dt}\right)^{\rm (pqs)} &=&
\frac{9}{2}\,\frac{\;M_m}{M}\,R^5\s(GM)^{-5/6}\s a^{-5/2}\s\frac{d}{dn} \left[\s n^{11/3}\s K_2(2n-2{\Omega}_p) \s\right]\\
\label{derivative}\\
\nonumber
&=& 9\frac{\,M_m}{M}\left(\frac{R}{a}\right)^5 n \left[ \frac{11}{6} K_2(2n-2{\Omega}_p) + n \left(1-\frac{d{\Omega}_p}{dn}  \right)\frac{dK_2(\omega)}{d\omega}  \s\right]\;,
\ea
where we introduced the shortened notation $\omega\equiv\omega_{2200}=2(n-{\Omega}_p)\s$, and used the equality
\ba
\frac{d}{da}\,=\,\frac{dn}{da}\,\frac{d}{dn}\,=\;-\;\frac{3}{2}\s(GM)^{1/2}\s a^{-5/2}\s \frac{d}{dn}\,\;.
\label{}
\ea

In the low-frequency limit, the behaviour of terrestrial bodies is not very different {from that of a body following the Maxwell rheology}. For a near-spherical Maxwell body assumed homogeneous, the quality function reads (see, e.g. the Appendix in \citeauthor{ME2014} \citeyear{ME2014}):
\ba
K_2(\omega)\,=\,\frac{3}{2}\,\frac{{\cal A}_2\, \omega\, \tau_{_M}}{1\,+\,(1\s+\s{\cal A}_2)^2\,(\omega\s\tau_{_M})^2}  \;\,.
\label{K2}
\ea
the dimensionless factor given by
\ba
{\cal A}_2\s=\,\frac{57}{8\s\pi}\;\frac{\mu}{G\s(\rho\s R)^2}\;\,.
\label{A2}
\ea
Here $R$, $\mu$, $\rho$, and $\tau_{_M}$ are the radius, mean unrelaxed rigidity, mean density, and the mean Maxwell time of the body, while $G$ is Newton's gravitational constant.

$K_2(\omega)$ is a kink-shaped odd function. Within the interpeak interval of values of $\omega$, this function is close to linear, its derivative $dK_2/d\omega$ staying positive. Therefore, within this interval, the sign of the second term on the right-hand side of equation (\ref{derivative}) coincides with the sign of the difference $\,1\s-\s d{\Omega}_p/dn\,$, where $\s{d\Omega_p}/{dn}\s$ is given by formula (\ref{3ksi.eq}).

In the point of resonance, the first term on the right-hand side of expression (\ref{derivative}) vanishes, and stability is determined by the second term only:
\ba
\frac{d}{da}
\left(\frac{da}{dt}\right)^{\rm (pqs)}_{
\,n={\Omega}_{p}
}=
9\frac{\,M_m}{M}\left(\frac{R}{a}\right)^5 n^2
\left(1-\frac{d{\Omega}_p}{dn}  \right)\frac{dK_2(\omega)}{d\omega}
\Bigg{|}_{\,\omega=0}
\ea
It therefore turns out that synchronism is stable if $\s{d\Omega_p}/{dn}>1\s$, and is unstable otherwise.

\bibliography{references}
\bibliographystyle{plainnat}

\end{document}